\documentclass[twocolumn,a4paper,superscriptaddress,showpacs,showkeys]{revtex4}
\usepackage[ansinew]{inputenc}
\usepackage[dvips]{graphicx}
\usepackage{psfig}
\usepackage{amssymb}
\begin{document}
\title{The Phase Information Associated to Synchronized Electronic Fireflies}

\author{J.L. Guisset}
\affiliation{Universit\'e Libre de Bruxelles, Belgium}
\author{J.L. Deneubourg}
\affiliation{Universit\'e Libre de Bruxelles, Belgium}
\author{G.M. Ram\'{\i}rez \'Avila}
\thanks{Partially supported by Belgium Technical Cooperation}
\affiliation{Universit\'e Libre de Bruxelles, Belgium}
\affiliation{Universidad Mayor de San Andr\'es, La Paz, Bolivia}

\begin{abstract}
An electronic implementation referring to fireflies ensembles
flashing in synchrony in a self-organization mode, shows the
details of the phase-locking mechanism and how the phases between
the electronic oscillators are related to their common period.
Quantitative measurements of the timing signals link the limits of
a steadily established synchronization to the physics of the
electronic circuit. Preliminary observations suggest the existence
of bifurcation-like phenomena.
\end{abstract}


\maketitle
\section{Introduction}
\label{intro} Self-organization is a widespread feature appearing
under a variety of living and inanimate systems. Synchronization
that can be understood as an adjustment of rhythms of oscillating
objects due to their weak interaction \cite{PIKOVSKY}, represents
one of the forms of self-organized matter \cite{BLEKHMAN}.

There are numerous examples of systems of coupled oscillators able
to induce structured behaviors between the interacting oscillators
\cite{MIROLLO-STROGATZ,STROGATZ,WINFUL,WIESENFELD,HOHL}.

The synchronized flashes of huge ensembles of fireflies in
south-asian countries swarm trees is one of these surprising
self-organization effects. The phenomenon was already mentioned
three centuries ago by the Dutch physician Kaempfer in 1727
\cite{BUCK-BUCK3}, but it is only recently that experimental (see
e.g.
\cite{BUCK,BUCK-BUCK1,BUCK-BUCK2,LLOYD1,LLOYD2,MOISEFF-COPELAND})
and theoretical \cite{ERMENTROUT,MIROLLO-STROGATZ} researches
suggested an adequate operational model \cite{CAMAZINE}.

At the individual firefly level, the rhythm of the recurring
flashes is supposed to be under the control of a neural center
which itself may be optically influenced by the flashes of
neighboring fireflies. From an experimental point of view, it is
clear that the fireflies interact and modify each other's rhythms,
which automatically leads to the acquisition of synchrony.

Although the anatomical details of the neural activity are largely
unknown, a model has been proposed which accounts for the
essential operational parameters. The model is based on a
relaxation oscillator in which it is possible to reset the duty
cycle by optical means. Moreover, the reset action is phase
dependent: the duty cycle is lengthened or shortened depending of
the time interval between the flashes of the interacting
fireflies.

By constructing an electronic implementation of it, Garver and
Moss showed that the model worked as it was supposed to do
\cite{GARVER-MOSS}. They report ensemble behaviors which indeed
are analogous to what is observed with fireflies, although they
experimented on a much smaller scale.

We constructed an ``open'' version of this electronic firefly,
whose free-run duty cycle can be modified and adjusted manually on
the spot, and on which quantitative measurements of periods and
phase differences may be performed with the required precision. We
call it ``LCO'' the acronym of Light Controlled Oscillator.

Compared to a firefly, the workings of an LCO are without any
mystery, which allows for a detailed quantitative description of
the synchronization mechanism at least for small ensembles.

Our aim is to experiment with LCO's and to investigate the local
level features of the self-organization they exhibit. At first we
are looking for the parameters involved in a steady
synchronization state achieved by two LCO's, bringing out the
factors leading to synchrony. It appears that the period of
synchronized LCO's is tightly related to the phase difference
between them, showing how the electronics of the synchronization
actually works, and why synchrony acquisition ceases outside
limits set by the physics of the system. Similar phases
relationships have been found for systems of three LCO's and more,
revealing interesting features.
\section{Presentation of an LCO}
\label{lco} 
\begin{figure}[hbtp!]
 \begin{center}
   \hbox{
       \includegraphics[scale=0.75]{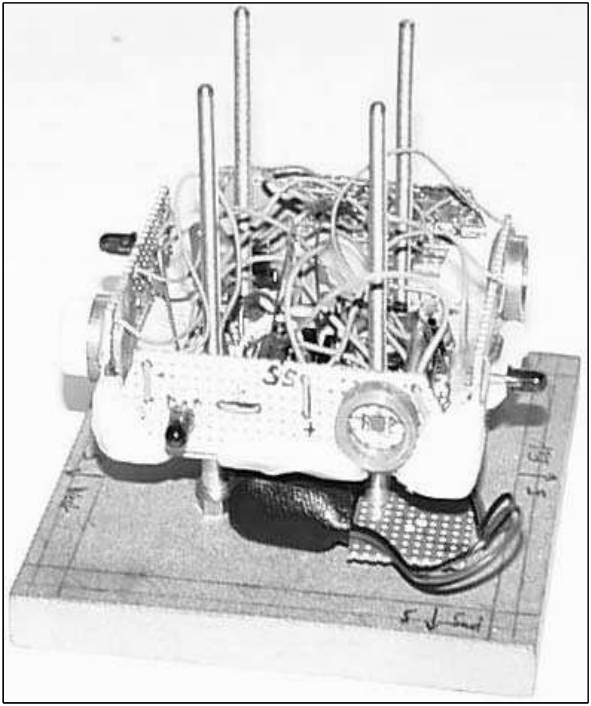}
       }
 \end{center}
    \vskip-.3in
  \caption{View of a single LCO.}
  \label{f1}
\end{figure}

Basically, our LCO (Fig. \ref{f1}) is composed of a
LM555 circuit wired as an astable \cite{LM555}, the alternations
of which are determined by a dual RC circuit in parallel with four
photo-sensors \cite{GARVER-MOSS}.

\begin{figure}[hbtp!]
 \begin{center}
   \hbox{
       \includegraphics[scale=0.5]{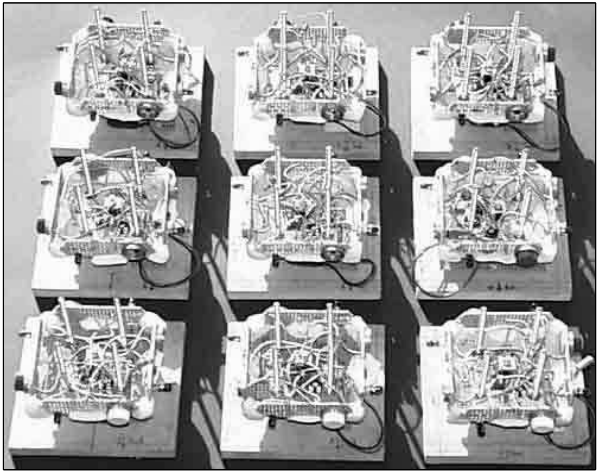}
       }
 \end{center}
    \vskip-.3in
  \caption{Group of nine LCO's.}
  \label{f2}
\end{figure}

We made nine LCO's (Fig. \ref{f2}): nine autonomous oscillators
coupled by their IR beams. They had much success when, disposed on
a table, they went to synchronize like exotic fireflies which they
are aimed to mimic. Each LCO is a module made of the same
electronic components and having the same structure. A square base
(11 cm $X$ 11 cm) gives the over-all horizontal dimensions of each
LCO in the global pattern. With nine LCO's it is possible to
achieve several different patterns.

Each base may sustain several printed circuits giving the
possibility of vertical extension, but keeping the same over-all
horizontal dimensions.

For the time being, our LCO's have two levels (Fig. \ref{f1}). The
lower part consists of a 9 volt battery and its clamping system.
The oscillator's printed circuit with the variable resistors
allowing the adjustment of the period's two time intervals, makes
the upper part of a LCO module. The circuit is square-like too but
its size is smaller than the basis. Even smaller printed circuits
bearing each an infrared LED and a photo-sensor, are fixed
vertically on the sides of the upper part. Provision is made to
mask the sensors allowing the LCO's to oscillate "in the dark". In
the aim of public presentations, the upper part bears a fifth LED
flashing visible light in synchrony with the IR one's, just to
produce a ``firefly effect''.

\begin{figure}[hbtp!]
 \begin{center}
   \hbox
       {
       \includegraphics[scale=0.5]{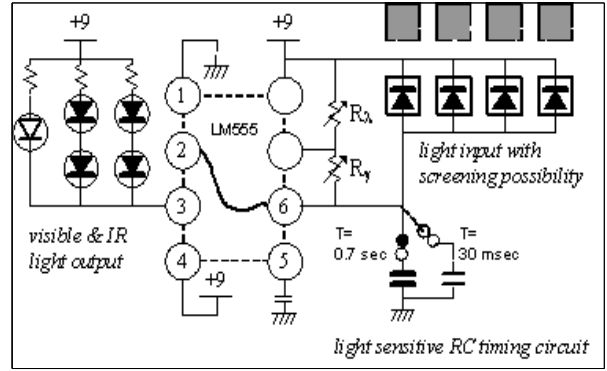}
       }
 \end{center}
      \vskip-.3in
  \caption{Block diagram of the  LCO.}
  \label{f3}
\end{figure}

The RC timing components of the LM555 consist of two resistors and
a single capacitor (Fig. \ref{f3}). Let $R_{\lambda}$,
$R_{\gamma}$ and $C$ be the values of those components responsible
for the LCO's timing with masked photo-sensors (timing ``in the
dark'').

The period is made up of two states: a longer one that may be
changed by manually adjusting $R_{\lambda}$, a shorter one that
may be changed by acting on $R_{\gamma}$. The LED's are wired at
the output of the LM555, switching on during the shorter part of
the period.

The photo-sensors act as current sources when they are receiving
light, shortening the charging time of the capacitor and making
longer the time required to discharge it.

In our model the resistors $R_{\lambda}$ and $R_{\gamma}$ are
partially variable
\[
  \begin{array}{lll}
     R_\lambda \, &=&\, 68 \, \text{k}\Omega  \, +\,[0, 50] \, \text{k}\Omega  \\
     R_\gamma \, &=&\, 1.2 \,\text{k}\Omega \, +\, [0.0, 1.0] \,
     \text{k}\Omega.
  \end{array}
\]

We use the same LCO's for two different missions: firstly, for
demonstrations where it is required to carry out synchronization
at a pace of about one flash per second (very impressive), and
secondly for observing the synchronization by an oscilloscope,
which requires a period of about 30 ms. This change of period
range is made by modifying nothing else but the capacitor value,
which has the advantage of leaving the lighting percentage per
period unchanged (less than 2\%) because the
$R_{\lambda}/R_{\gamma}$ ratio is not modified:
\[
  \begin{array}{lll}
  C \ =\, 10 \ \mu \text{F} & \rightarrow & {\mathrm period} \ \ T \ =\  0.7 \ \text{s} \\
  C \ =\, 0.47 \ \mu \text{F} & \rightarrow & {\mathrm period} \ \ T \ =\ 30 \
  \text{ms}
  \end{array}
\]

As the illumination of the photo-sensors modifies the period of an
LCO, it is useful to distinguish the functioning of it in the dark
from its functioning when receiving light pulses from its
neighbors or diffuse light from the surroundings.

When all the photo-sensors of an LCO are masked, its period
depends only from its electronics. We took this particular period
as a reference for each LCO.
\begin{figure}[hbtp!]
 \begin{center}
   \hbox
       {
       \includegraphics[scale=0.5]{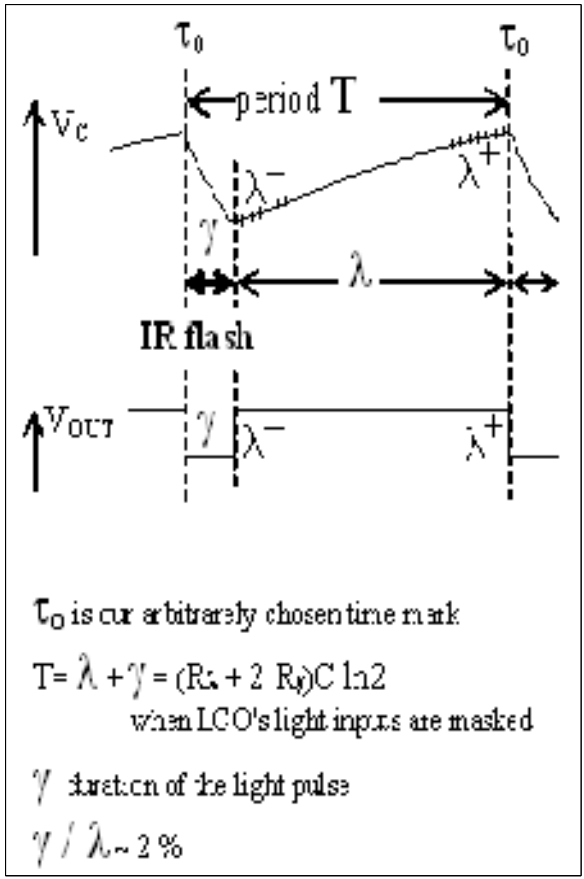}
       }
 \end{center}
      \vskip-.3in
  \caption{Definition of parameters in an LCO.}
  \label{f4}
\end{figure}

In the framework of this article, we use the following parameters
related to the synchronization of the LCO (Fig. \ref{f4}):

\begin{itemize}
\item $\lambda=(R_{\lambda}+R_{\gamma})C\ln 2$, the larger
alternation time, it corresponds with a capacitor's charging time
between $1/3$ and $2/3$ of the total charge. \item
$\gamma=R_{\gamma}C\ln 2$, the shorter alternation time, it
corresponds with a capacitor's discharging time within the same
limits. \item $\lambda^-, \lambda^+$, the beginning and the end of
a long alternation. \item $\tau^o$, the instant coinciding with
the transition from $\lambda$ to $\gamma$ ; here after we will use
this parameter as the ``reference moment'' in the period of an
LCO. \item $T_s$, the common period of a set of synchronized
LCO's. \item $T_A, \tau^o_A, \lambda_A, \gamma_A, T_B, \tau^o_B,
\lambda_B, \gamma_B, \ldots$, the periods, the ``reference
moments'', and the duration of the alternations of LCO$_A$,
LCO$_B$, $\ldots$ in illumination situations. \item $T^d_A, T^d_B,
T^{\text{dark}}_C,\ldots, \gamma^{\text{dark}}_A,
\gamma^{\text{dark}}_B, \gamma^{\text{dark}}_C, \ldots$, the
periods and the durations of the short and long alternations ``in
the dark'' of LCO$_A$, LCO$_B$, LCO$_C$, $\ldots$ i.e. when all
their photo-sensors are masked.
\end{itemize}
\section{An LCO coupled to a short pulse blind LCO}
\label{bf} In order to investigate the mechanisms inducing the
synchronization, we have placed an LCO, namely LCO$_B$, in
interaction with the equivalent of an LCO ``kept in the dark''.

Let LCO$_A$ be this blind LCO; its IR pulse (lasting during
$\gamma_A^{\text{dark}}$) has been reduced to a quarter of
$\gamma_B$. In practice it is obtained from a low frequency
generator controlling a monostable producing a $\gamma_A$ pulse of
constant duration and sufficiently short.

The signals are picked up at the low impedance output of the
LM555; the measurements of phases and periods are carried out with
an oscilloscope {\em Tektronix TDS 3012} according to well-known
procedures; observations are done without difficulty with a
precision of about 0,1\%. To be coherent with the measurements
presented further on in this article, the triggering is provided
by the LCO$_A$ considered as the reference LCO.
\begin{figure}[hbtp!]
  \begin{center}
  \hbox{
        \includegraphics[scale=0.45]{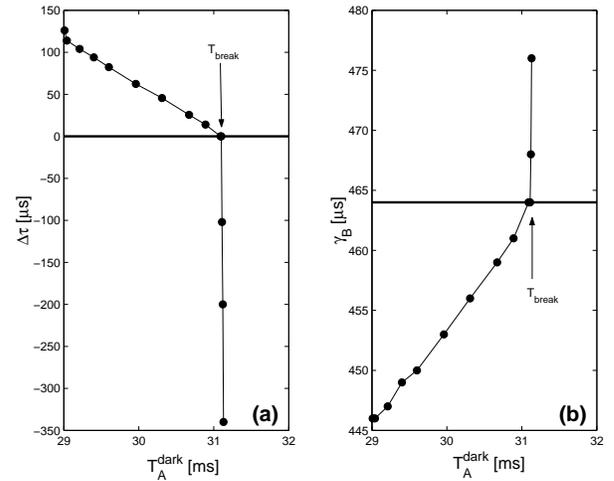}
     }
    \end{center}
     \vskip-.3in
  \caption{(a) Time difference as a function of the short
  impulsion oscillator period. (b) Duration of the LCO$_A$'s short
  alternation as a function of the short impulsion oscillator period.}
  \label{f5}
\end{figure}

From the very first observation it is obvious that the
synchronization implies a phase relation between the two
oscillators.

Fig. \ref{f5}(a) represents $\Delta \tau=\tau^o_B-\tau^o_A$ (i.e.
the position of $\tau^o_B$ related to $\tau^o_A$, the latter being
taken as reference) as a function of the period
$T_A^{\text{dark}}(=T_s)$ of the blind LCO$_A$.

\begin{figure}[hbtp!]
  \begin{center}
   \hbox{
        \includegraphics[scale=0.7]{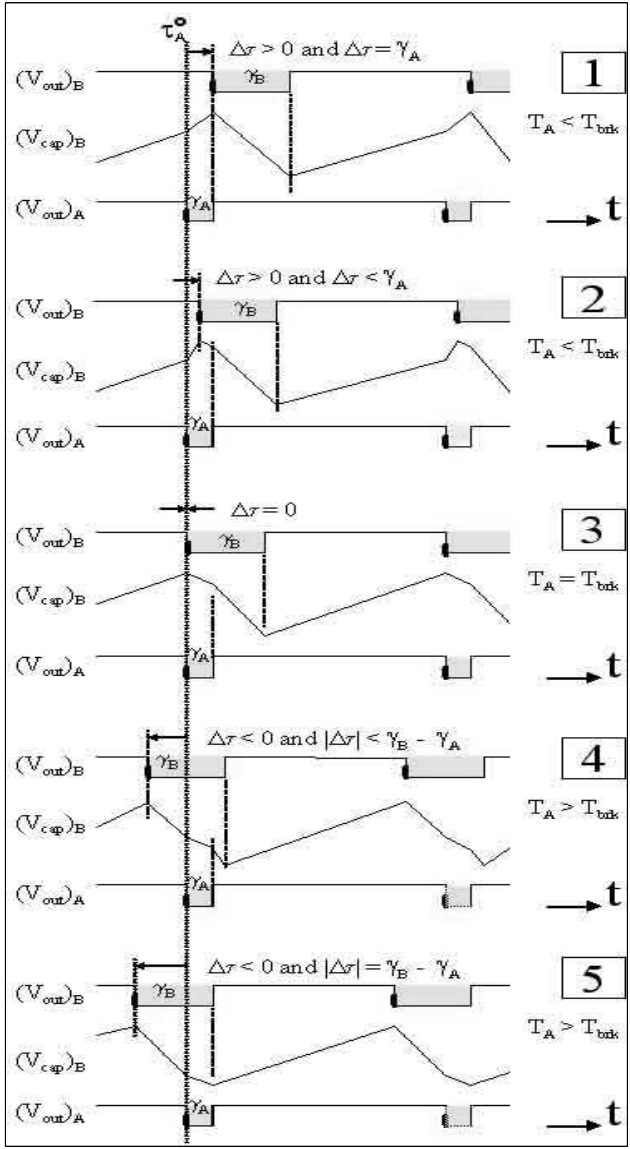}
     }
  \end{center}
     \vskip-.3in
  \caption{Signal shape for several phase differences between LCO$_B$
   and the short pulse blind LCO$_A$.}
  \label{f6}
\end{figure}

When $\Delta\tau >0$, i.e. when $\tau^o_A$ appears before
$\tau^o_B$ (Fig. \ref{f6} cases 1 and 2), the phase-control is
stable and $T_s$ can be measured easily.  However, for
$\Delta\tau<0$ (Fig. \ref{f6} cases 4 and 5), the stability is
much more precarious, even if also in that situation there is a
synchronization of LCO$_B$ to the blind LCO$_A$.

The shape of $\Delta \tau$ as a function of
$T_A^{\text{dark}}(=T_s)$ can be fairly well schematized by two
straight lines intersecting at $\Delta\tau=0$ for the abscissa
$T_{\text{break}}$.

The plot suggests two phase-control modes, situated at either side
of $\Delta\tau=0$ (Fig. \ref{f6}, case 3), distinguishing
themselves by two parameters which are easy to measure:
\begin{itemize}
\item[-] The domain of the phase-locking, that is to say the
interval limited by the periods for which there is
synchronization. \item[-] The slope of the straight lines, which
represents somehow the gain of a servo-system's feedback control.
\end{itemize}
These two modes correspond to different control mechanisms:
\begin{itemize}
\item For $\Delta \tau<0$, there can be only widening of the
period because the illumination and thus the photo-current are
totally included in the interval $\gamma_B$ (Fig. \ref{f6} cases 4
and 5), under these circumstances the photo-current adds to the
discharging current of the capacitor $C$ through $R_{\gamma}$. As
the extension of $\gamma$ corresponds with an increase of the
period, the phase-control is stable. Nevertheless the influence of
the photo-current on the extension of the period is of little
importance, first because the discharge current is by two orders
of magnitude superior to the charge current, and secondly because
$\gamma$ represents only 2\% of the period. The upper limit of the
phase-locking is reached as soon as the photo-current shortens the
alternation $\lambda_B$ of the following period, that is to say as
soon as the positive feedback resulting from this situation makes
the phase-control unstable. \item For $\Delta \tau>0$, Only a
shortening of the free-run period $T_B$ is possible. The
shortening takes place during $\lambda^+$ the end of the interval
$\lambda$, by increasing the charging speed of capacitor $C$ due
to the photo-current brought in, in parallel with $R_{\lambda}$.
Due to the photo-current, the voltage of $C$ reaches more rapidly
the value $V_C=2V_M/3$ which triggers the switching from the
$\lambda$ alternation towards the $\gamma$ alternation. The lower
limit of the phase-locking domain is reached when the duration
$(\Delta \tau)_{\text{min}}=\gamma_A$ (Fig. \ref{f6}, case 1)
during which the photo-current speeds up the charging of the
capacitor, is not sufficient anymore to reach the switching point
of $\lambda$ towards $\gamma$. As a consequence, for the shortest
periods, the synchronization depends strongly on the intensity of
the light received by phase-locked LCO$_B$. \item For
$0<\Delta\tau<\gamma_A$, This constitutes the intermediate
situations (Fig. \ref{f6}, case 2): the excess of photo-current
acts in accordance with the mode $\Delta\tau<0$ already described,
achieving a not so important lengthening of the synchronized
period as compared to its shortening.
\end{itemize}
\section{An LCO coupled to a blind LCO flashing large pulses}
\label{bfl} 
\begin{figure}[hbtp!]
  \begin{center}
  \hbox{
     \includegraphics[scale=0.45]{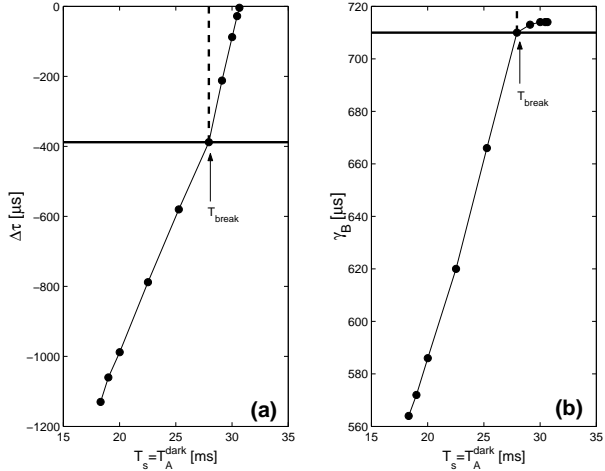}
       }
   \end{center}
     \vskip-.3in
  \caption{(a) Time difference as a function of the long impulsion
  blind LCO period. (b) Duration of the LCO$_B$'s short alternation
  as a function of the long impulsion blind LCO period.}
  \label{f7}
\end{figure}

The experimental setup that has been used to get Fig.
\ref{f7}, differs from the previous one, only by the width of the
flashes emitted by the blind LCO: they have been widened to
increase the ratio $\gamma_A/\gamma_B$ from $1/4$ to 2, allowing
the blind LCO flashes to overlap the synchronized LCO flashes.

The observation of this second type of synchronization is
important because it looks more like that of an actual LCO pair in
mutual interaction, for which differences between the $\gamma$
alternations are the rule with occurrences of overlapping.

\begin{figure}[hbtp!]
  \begin{center}
  \hbox{
        \includegraphics[scale=0.62]{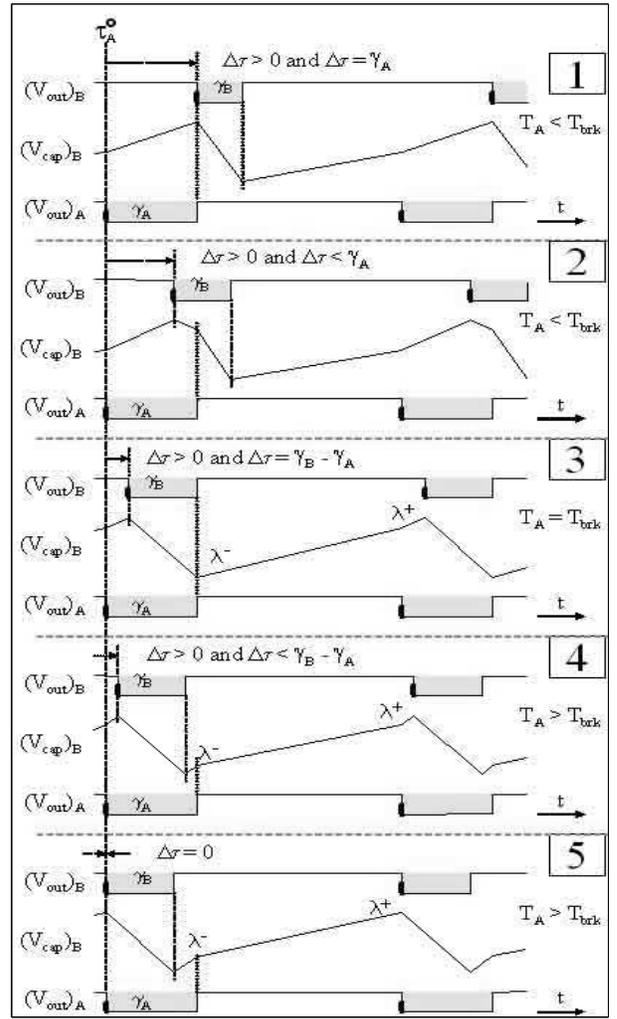}
     }
   \end{center}
     \vskip-.3in
  \caption{Signal shape under different conditions for the coupling
  of a LCO$_B$ with a long pulse blind LCO$_A$.}
  \label{f8}
\end{figure}

In the phase-locking with short pulses ($1/4$ ratio) one
distinguishes five unambiguous situations of synchronization.
However, when the alternation  $\gamma_A$ is significantly larger
than $\gamma_B$, it is possible that it overrides $\gamma_B$ by
illuminating the end $\lambda^+$ and the beginning $\lambda^-$ of
the neighboring $\lambda_B$ intervals of  $\gamma_B$ (Fig.
\ref{f8}, case 4).

The graphs $\Delta \tau$ and $\gamma_B$ as a function of
$T_s=T_A^{\text{dark}}$ in Figs. \ref{f7}(a) and (b), show a
singular value $T_A^{\text{dark}}=T_{\text{break}}$ for which
there is a change of slope.
$T_{\text{break}}$ corresponds to a situation in which $\gamma_A$
covers $\gamma_B$ completely and is about to start the covering of
$\lambda^-_B$ (the start of the following interval $\lambda_B$
(Fig. \ref{f8}, case 3).

We observe that :
\begin{itemize}
\item[a)] For $T_A<T_{\text{break}}$, the interval $\gamma_A$
covers over two alternations of LCO$_B$ (Fig. \ref{f8}, case 2);
the phase-control acts as in the case $\gamma_A \ll \gamma_B$
(Fig. \ref{f6}, case 2). In the same way, the lower limit of the
synchronization domain depends of the total width $\gamma_A$.
\item[b)] For $T_A>T_{\text{break}}$, the interval $\gamma_A$
covers three alternations of LCO$_B$ (Fig. \ref{f8}, case 4) and
the phase-control functions differently:
\begin{description}
\item[In Fig. \ref{f7}(a)]: $\gamma_B$ is constant because it
remains entirely covered by $\gamma_A$; however $\gamma_A$ is
sufficiently large to illuminate either sides of $\gamma_B$, that
is to say to ensure the phase-control by truncating $\lambda^+$
(the end of alternation $\lambda_B$) while at the same time
shortening $\lambda^-$) the beginning of the following
$\lambda_B$.
\item[In Fig. \ref{f7}(b)]: $T_s=T_A^{\text{dark}}$
varies more slowly indicating a change of phase-control mode,
indeed the increase of the period results from the combination of
a constant extension of $\gamma_B$ and of a shortening of the two
neighboring alternations $\lambda^-_B$ and $\lambda^+_B$.
\end{description}
\end{itemize}

Finally, Fig. \ref{f7}(a) shows that the upper limit of the
synchronization domain is reached when $\Delta\tau=0$ (Fig.
\ref{f8}, case 5), i.e. when the time-marks $\tau^o_A$ and
$\tau^o_B$ are superposed. This observation is of major importance
for the analysis of the mutual synchronization between two
interacting LCO's , neither being blind. Indeed the loss of
synchronization at $\Delta\tau=0$, means that shortening
$\lambda^-_B$ at the start of a alternation is not sufficient to
maintain the phase-control, the latter working only when it is the
end of $\lambda$ which is truncated, i.e. $\lambda^+_B$.
\section{Synchronization between two interacting LCO's. Measure and analysis}
\label{2int} 
\begin{figure}[hbtp!]
  \begin{center}
  \hbox{
        \includegraphics[scale=0.6]{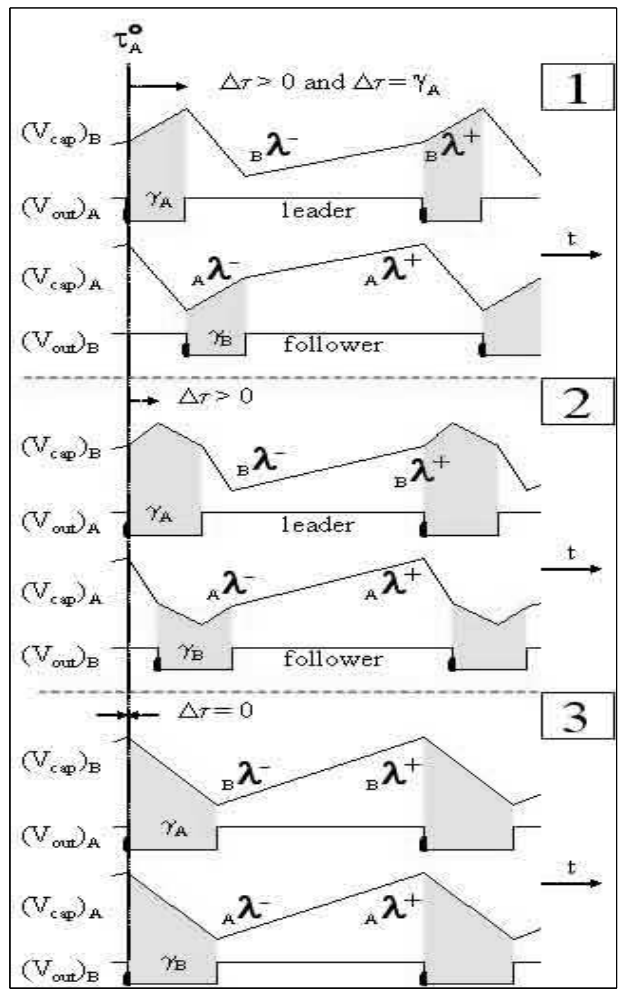}
     }
   \end{center}
     \vskip-.3in
  \caption{Signal shape under different conditions for two coupled
   LCO's.}
  \label{f10}
\end{figure}

When two interacting LCO's synchronize, their short
alternations $\gamma_A$ and $\gamma_B$ cover each other mutually,
including their time-mark $\tau^o_A$ and $\tau^o_B$. As a
consequence, the fractions of the alternations $\gamma_A$ and
$\gamma_B$ which are not superposed illuminate the preceding
$\lambda^+$ and the following $\lambda^-$ (Fig. \ref{f10}, case 2
and its symmetrical case which is not represented).

This situation is similar to that described at the end of the
preceding paragraph (Fig. \ref{f8}, case 4 and 5); it allows us to
deduce that of the two parts, $\lambda^+$ and $\lambda^-$ of an
alternation $\lambda$, it is only $\lambda^+$ that controls the
synchronization in a decisive way. Indeed (Fig. \ref{f8}, case 5)
shows that phase-locking and synchronization stops as soon as
$\lambda^+$ ceases to be illuminated.

\begin{figure}[hbtp!]
  \begin{center}
   \hbox{
        \includegraphics[scale=0.45]{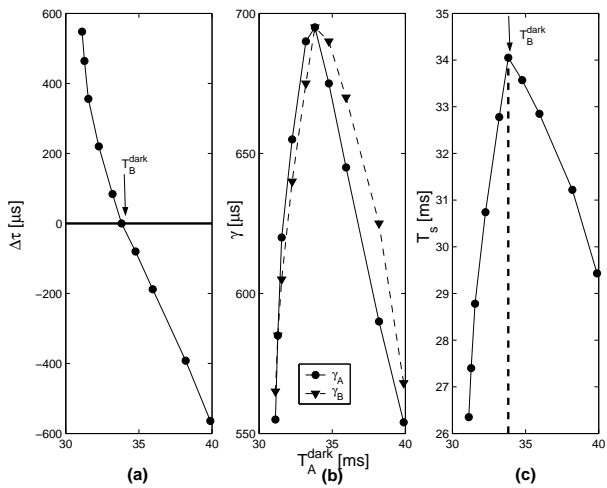}
        }
  \end{center}
     \vskip-.35in
  \caption{Variation of some magnitudes as a function of the LCO$_A$
  period in the dark. (a) Time difference. (b) Short alternations
  $\gamma_A$ y $\gamma_B$ of both LCO's. (c) Synchronization
  period.}
  \label{f9}
\end{figure}

The plots of $\Delta\tau$, $\gamma_A$ and $\gamma_B$, and $T_s$,
as a function of $T_A^{\text{dark}}$ in Figs. \ref{f9}(a), (b) and
(c) show the binary structure of the interaction between two
LCO's: In Fig. \ref{f9}(a), may be of two polarities in an
equivalent way, indicating that the two LCO's are interchangeable.
In Fig. \ref{f9}(b), the time intervals $\gamma_A$ and $\gamma_B$
change in the same manner when they are overlapping. When
$\gamma_A$ and $\gamma_B$ are exactly superimposed, i.e. when
$\Delta\tau=0$ (Fig. \ref{f10}, case 3), their length is at a
maximum as does $T_s$ the common period of the synchronized LCO's
(Fig. \ref{f9}(c)).

On both sides of the maximum of $T_s$, the light associated with
the one or the other alternation $\gamma$ truncates the
corresponding $\lambda^+$ (Fig. \ref{f10}, case 1, 2, 4 and 5). As
observed earlier, the photo-effect on $\lambda^-$ is not as strong
as on $\lambda^+$. In fact two interacting LCO's have not the same
status: one is a leader who cuts the $\lambda^+$ part of his
coupled partner, the latter being the follower who cuts the
$\lambda^-$ part of his leader. However this status may change,
depending of the sign of $\Delta\tau$, i.e. the relative positions
of $\tau^o_A$ and $\tau^o_B$.
\section{Synchronization between several LCO's}
\label{3int} As long as the interacting system has a binary
symmetry, the choice of the reference LCO is irrelevant and is
without importance for the quantitative observation of the
synchronization: the reference LCO is simply that one whose period
$T_A^{\text{dark}}$ is modified manually. On the opposite, for
sets of more than two LCO's  it is mandatory to specify the
position of the reference LCO in that of the interacting LCO's.

We have been able to observe phase-locking in sets made up of 5
LCO's, using two {\em Tektronix TDS 3012} oscilloscopes triggered
simultaneously by the output signal of the reference LCO. However
those last measurements were rather difficult to perform, due to
the intrinsic instability of the LM555 oscillators and/or the
presence of multiple states.
\begin{figure}
  \begin{center}
  \hbox{
     \includegraphics[scale=0.45]{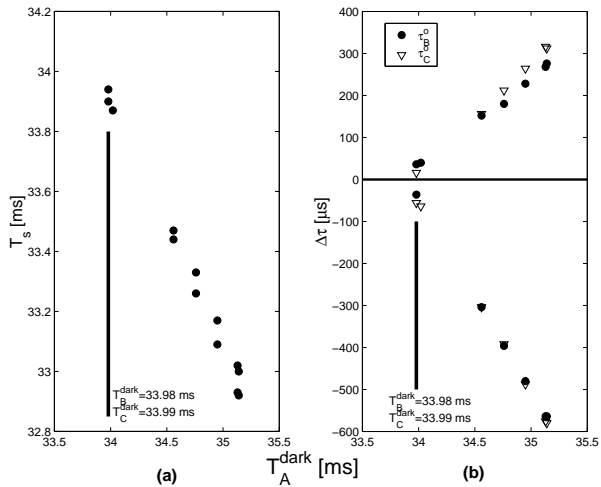}
    }
   \end{center}
     \vskip-.3in
  \caption{Variation of some magnitudes as a function of the
  LCO$_A$ (reference, in the mid position) period in the dark in
  the case of three LCO's interacting in line.(a) Synchronization
  period. (b) Time difference of LCO$_B$ and LCO$_C$ with respect
  to LCO$_A$.}
  \label{f11}
\end{figure}

Fig. \ref{f11}(a) and (b) have been obtained with a set of 3 LCO's
put in line: LCO$_B$--LCO$_A$--LCO$_C$. The reference was
LCO$_A^{\text{dark}}$ in the mid position. Synchronization has
been achieved for $T_A^{\text{dark}}$ varying between 33.98 ms and
35.2 ms. The ``periods in darkness'' of the other are:
$T_B^{\text{dark}}$=33.98 ms and $T_C^{\text{dark}}$=33.99 ms.

Fig. \ref{f11}(a) shows a bifurcation-like phenomenon: the 3-LCO
system synchronizes according to two modes brought out by two
significantly different values of $T_s$ as a function of
$T_A^{\text{dark}}$.

Fig. \ref{f11}(b) shows the same behavior by plotting the
time-marks $\tau^o_B$ and $\tau^o_C$ as a function of
$T_A^{\text{dark}}$, taking $\tau^o_A$ as the reference time-mark.

There is obviously a symmetry in this 3-LCO synchronization: once
the LCO's are synchronized, the phases may equally well have one
or the other polarity (Fig. \ref{f11}(b)). Moreover, a closer look
at the data producing Fig. \ref{f11}(b) shows that
$\Delta\tau^o_B$ and $\Delta\tau^o_C$ are always of opposite
polarity; this suggests that LCO$_B$ and LCO$_C$ are in some way
interchangeable. However, before the synchronization has been
settled, it is not possible to foresee the mutually exclusive
polarities of $\tau_B$ and $\tau_C$, confirming that a
bifurcation-like phenomenon is present.
\section{Conclusion}
\label{conc} Obviously, synchronization is tightly linked to
phase-locking and the domain associated with it. This is why we
considered as a criterion that coupled LCO's are synchronized only
if they exhibit a stable dependency of their phase as a function
of their period differences.

Our measurements show that LCO's synchronize in as much that the
phases involved in the phase-locking feedback do not exceed values
related to the widths of the coupling light pulses.

Finally we suggest using the criterion associating synchronization
and phase-locking domain as a test to compare theoretical models
of interacting LCO's to their experimental implementations. In a further
publication, we analyze this situation.


\begin{thebibliography}{10}

\bibitem{PIKOVSKY}
A. Pikovsky, M. Rosenblum and J. Kurths, {\em Synchronization a
universal concept in nonlinear sciences\/} (Cambridge University
Press, Cambridge, 2001).

\bibitem{BLEKHMAN}
I. Blekhman, {\em Synchronization in science and technology\/}
(Asme Press, New York, 1988).

\bibitem{MIROLLO-STROGATZ}
Mirollo RE, Strogatz SH, Synchronization of pulse-coupled
biological oscillators , {\em SIAM Journal of Applied Mathematics}
{\bf 50} (1990) 1645-1662.

\bibitem{STROGATZ}
S.H. Strogatz, R.E.Mirollo and P.C. Matthews, Coupled nonlinear
oscillators below the synchronization threshold: Relaxation by
generalized Landau damping, {\em Phys. Rev. Lett.\/} {\bf 68}
(1992) 2730--2733.

\bibitem{WINFUL}
H.G. Winful and L. Rahman, Synchronized chaos and spatiotemporal
chaos in arrays of coupled lasers, {\em Phys. Rev. Lett.\/}{\bf
 65} (1990) 1575--1578.

\bibitem{WIESENFELD}
K. Wiesenfeld, P. Colet and S.H. Strogatz, Synchronization
transitions in a disordered Josephson series array, {\em Phys.
Rev. Lett.\/} {\bf 76} (1996) 404--407.

\bibitem{HOHL}
A. Hohl, A. Gavrielides, T. Erneux and V. Kovanis, Localized
synchronizationin two coupled nonidentical semiconductor lasers,
{\em Phys. Rev. Lett.\/} {\bf 78} (1997) 4745--4748.

\bibitem{BUCK-BUCK3}
J. Buck and E. Buck, Mechanism of rhythmic synchronous flashing of
fireflies, {\em Science\/} {\bf 159} (1968) 1319--1327.

\bibitem{BUCK}
J. Buck, E. Buck, J.F. Case and F.E. Hanson, Control of flashing
in fireflies. V. Pacemaker synchronization in \textit{Pteroptyx
cribellata}, {\em J. Comp. Physiol.\/} {\bf A 144} (1981)
287--298.

\bibitem{BUCK-BUCK1}
J. Buck and E. Buck, Synchronous fireflies, {\em Sc. Am.\/} {\bf
234} (1976) 74--85.

\bibitem {BUCK-BUCK2}
J. Buck, E. Buck, Toward a functional interpretation of
synchronous flashing by fireflies, {\em Am. Naturalist.\/} {\bf
112} (1978) 471--492.

\bibitem {LLOYD1}
J.E. Lloyd, Fireflies of Melanesia: Bioluminiscence, mating
behavior and synchronous flashing (\textit{Coleoptera:
Lampyridae}), {\em Env. Entom.\/} {\bf 2} (1973) 991--1008.

\bibitem {LLOYD2}
J.E. Lloyd, Model for the mating protocol of synchronously
flashing fireflies, {\em Nature\/} {\bf 245} (1973) 268--270.

\bibitem{MOISEFF-COPELAND}
A. Moiseff and J. Copeland, A new type of synchronized flashing in
a North America firefly, {\em J. Ins. Behavior\/} {\bf 13} (2000)
597--612.

\bibitem{ERMENTROUT}
G.B. Ermentrout, An adaptive model for synchrony in the firefly
\textit{Pteroptyx malaccae} {\em J. Math. Biol.\/} {\bf 29} (1991)
571--585.

\bibitem{CAMAZINE}
S. Camazine, J.L. Deneubourg, S. Franks, J. Sneyd, G. Theraulaz
and E. Bonabeau {\em Self-organization in biological systems\/}
(Princeton University Press, Princeton, 2001).

\bibitem{GARVER-MOSS}
Garver W, Moss F, Electronic fireflies, {\em Sc. Am.\/} {\bf 269}
(1993) 128-130.

\bibitem{LM555}
LINEAR Databook, (1982), {\em LM555/LM555C Timer}. National
Semiconductor Corporation (9): 33-38.

\end{thebibliography}
\end{document}